# Characterization of the Soluble Nanoparticles Formed through Coulombic Interaction of Bovine Serum Albumin with Anionic Graft Copolymers at Low pH

Evaggelia Serefoglou[a], Julian Oberdisse[b], and Georgios Staikos[a*]

[a]Department of Chemical Engineering, University of Patras, GR-265 04 Patras, Greece and Institute of Chemical Engineering and High Temperature Chemical Processes, FORTH/ICE-HT, P.O. Box 1414, GR-26504 Patras, Greece

[b] Laboratoire des Colloïdes, Verres et Nanomatériaux, Université Montpellier II, France and Laboratoire Léon Brillouin CEA/CNRS, CEA Saclay, 91191 Gif sur Yvette, France

[*]To whom correspondence should be addressed. Fax: 30-2610-997-266. E-mail: **staikos@chemeng.upatras.gr**

# ABSTRACT

A static light scattering (SLS) study of bovine serum albumin (BSA) mixtures with two anionic graft copolymers of poly (sodium acrylate-*co*-sodium 2-acrylamido-2-methyl-1-propanesulphonate)-*graft*-poly (N, N-dimethylacrylamide), with a high composition in poly (N, N-dimethylacrylamide) (PDMAM) side chains, revealed the formation of oppositely charged complexes, at pH lower than 4.9, the isoelectric point of BSA. The core-corona nanoparticles formed at pH = 3.00, were characterized. Their molecular weight and radius of gyration were determined by SLS, while their hydrodynamic radius was determined by dynamic light scattering. Small angle neutron scattering measurements were used to determine the radius of the insoluble complexes, comprising the core of the particles. The values obtained indicated that their size and aggregation number of the nanoparticles, were smaller when the content of the graft copolymers in neutral PDMAM side chains was higher. Such particles should be interesting drug delivery candidates, if the gastrointestinal tract was to be used.

**Introduction**

The interactions of globular proteins with polyelectrolytes have attracted much interest.[1] What mostly occurs is the formation of insoluble complexes between oppositely charged entities, i.e. with polyanions at pH lower than the isoelectric point (pI) of the protein,[2-4] and with polycations at pH higher than pI.[5-7] This complexation can be used for protein separation.[8, 9] Nevertheless, the formation of soluble complexes may also occur in a relatively narrow pH range, just before the phase separation, [2, 10-12] as well as in protein-polyelectrolyte mixtures with an excess in the polyelectrolyte.[13]

Polyelectrolytes of a comb-type structure have been proposed for the preparation of water-soluble complexes through Coulombic interactions between polyelectrolytes and oppositely charged biological macromolecules, like DNA.[14-16] By grafting chains of a neutral hydrophilic polymer, like poly(ethylene glycol) (PEG), onto a cationic polymer backbone the formation of fully neutralized complexes with an oppositely charged polyelectrolyte, stabilized by means of the neutral PEG side chains is possible.

We have recently synthesised a similar comb-type copolymer, by grafting neutral poly (N,N-dimethylacrylamide) (PDMAM) chains onto an anionic poly(sodium acrylate-*co*-sodium 2-acrylamido-2-methylpropane sulphonate) backbone, [17] and studied the formation of water-soluble stoichiometric polyelectrolyte complexes with a polycation, namely poly (diallyldimethylammonium chloride).[18] We have moreover found that, if the graft copolymer contains more than 50 wt.% of PDMAM side chains, it forms also soluble complexes with bovine serum albumin (BSA), at pH lower than 4.9, i.e. the pI of BSA, where the protein molecules are positively charged.[19] It is expected such copolymers to be proved useful in oral drug delivery systems and for protein solubilization purposes.

In the present work, we have further investigated the above BSA complexation and characterized the complexes formed at pH = 3.0 between BSA and two such anionic graft copolymers containing 75 and 87.5 wt.% of PDMAM, by static and dynamic light scattering, as well as by small angle neutron scattering measurements.
.

**Experimental Section**

**Materials.** Bovine serum albumin (BSA) used was purchased from Sigma (A-7638) and used without further purification. Its molecular weight was determined by light scattering in 0.1 M NaCl and was found equal to 67 kDa.

The monomers, acrylic acid (AA), 2-acrylamido-2-methyl-1-propanesulphonic acid (AMPSA), N,N-dimethylacrylamide (DMAM) were purchased from Aldrich. Ammonium persulphate (APS, Serva), potassium metabisulphite (KBS, Aldrich), 2-aminoethanethiol hydrochloride (AET, Aldrich) and 1-(3-(dimethylamino) propyl)-3-ethyl-carbodiimide hydrochloride (EDC, Aldrich) were used for the synthesis of the graft copolymers.

**Polymer Synthesis and Characterisation.** Amine-terminated PDMAM was synthesised by free radical polymerisation of DMAM in water at 30 ºC for 6 h using the redox couple APS and AET as initiator and chain transfer agent, respectively. The polymer was purified by dialysis against water in a tubing with MWCO 12000 (Sigma) and then freeze-dried. Its number average molecular weight was determined by an acid-base titration of the amine end groups and was found equal to 15.000 Da.

A copolymer of AA and AMPSA, was prepared by free radical copolymerisation of the two monomers dissolved in water in a 1:5 mole ratio respectively, after a partial neutralisation (~90% mol) with NaOH at pH ~ 5-6, at 30 ºC for 6 h, using the redox couple APS/KBS. The product was obtained in its sodium salt form, poly(sodium acrylate-*co*-sodium 2-acrylamido-2-methyl-1-propanesulphonate) (P(NaA-*co*-NaAMPS)), after being fully neutralised with an excess of NaOH (pH~11), ultrafiltrated and freeze-dried. Its composition was determined by an acid-base titration and it was found to contain 18 mol % of acrylate units. Its molecular weight was determined by light scattering in 0.1 M NaCl and was found equal to $2.5 \times 10^5$ Da. It is worthy to point out that AMPSA units provide the copolymer with a negative charge, even at low pH, as they correspond to a strong acid, sulphuric acid, while AA units are necessary for the grafting reaction, that follows.

The graft copolymers P(NaA-*co*-NaAMPS)-*g*-PDMAMx, shortly designated as Gx, where x is the weight percentage feed composition in PDMAM, were synthesised through a coupling reaction between the AA units of the P(NaA-*co*-NaAMPS) copolymer and the amine-functionalised PDMAM. In a 5% aqueous solution of the polymer mixture, a fivefold excess of the coupling agent, EDC, was added and let under stirring for 12 h at room temperature. Addition of EDC was repeated for a second time. The EDC excess was subsequently fully neutralized with a fivefold quantity of NaOH and the products obtained were then purified by water with the Pellicon system and freeze-dried. The completion of the grafting reaction was confirmed by size exclusion chromatography, by means of a Waters system equipped with two Shodex OH-pak columns, B804 and B805. A schematic depiction of the graft copolymers is presented in Scheme 1.

Table 1 summarises the two graft copolymers, G75 and G87.5, synthesised and used in the present study. Their molar mass was determined by static light scattering measurements in 0.1 M NaCl.

**Static and Dynamic light scattering (SLS and DLS).** SLS and DLS measurements were conducted by means of a Model MM1 SM 200 spectrometer (Amtec, France) equipped with a real time granulometry (RTG) correlator (SEMATech, France). An He-Ne 10mW laser operating at 633 nm was used as a light source and either the light scattered at 90° was collected or complete series of measurements at different angles and concentrations were conducted for molecular weight determinations. The samples used, filtered through a 5 μ Millipore filter, were dust-free and optically transparent. The solutions used for molecular weight determination were moreover centrifuged for two hours at 15.000 turns per minute. The refractive index increments, *dn/dC*, values were measured by means of a Chromatix KMX 16 differential refractometer operating also at 633 nm. The results obtained were subjected to a Zimm analysis, and the molecular weight of the complexes was determined as the average of the values found by extrapolation to zero angle and zero concentration.

The DLS measurements were based on the calculation of a digital autocorrelation function $G_2(\tau)$, by the RTG programme and the correlator, so that the relaxation time of the correlation function, $1/\Gamma$, was determined. $\Gamma$ is given by $\Gamma = Dq^2$, where $D$ is the translation diffusion coefficient and $q$ the wave vector, defined as $q = (4\pi n/\lambda_o)\sin(\theta/2)$, $\theta$ being the scattering angle, $\lambda_o$ the laser wavelength and n the refractive index of the solvent. The hydrodynamic radius, $R_H$, was calculated using the Stokes–Einstein equation, $R_H = K_B T/6\pi\eta D$, where $K_B$ is the Boltzmann constant, T the absolute temperature and η the viscosity of the solvent.

**ζ-Potential.** ζ-Potential measurements were performed at 25 °C using a Zetasiser Nano ZS zeta potential analyzer (Malvern instruments).

**Small-angle neutron scattering (SANS).** SANS measurements were carried out at the Laboratoire Léon Brillouin (Saclay, France). The data were collected on beam line PACE at three configurations (6 Å, sample-to-detector distance 1 m; 7 Å and 18 Å, 4.55 m), covering a broad q range from 0.0023 to 0.32 Å$^{-1}$. 2 mm light path quartz cells were used. Empty cell scattering was subtracted and the detector was calibrated with 1 mm $H_2O$ scattering. All measurements were carried out at room temperature. Data were converted to absolute intensity through a direct beam measurement, and the incoherent background was determined with $H_2O/D_2O$ mixtures.

**Complex density.** The BSA/P(NaA-*co*-NaAMPS), protein-polyelectrolyte complex, was obtained as a precipitate, after mixing equal volumes of a BSA 1.00 x 10$^{-2}$ g/cm$^3$ with a P(NaA-*co*-NaAMPS) 2.50 x 10$^{-3}$ g/cm$^3$ aqueous solution, pH = 3.00, and dried in high vacuum for two days. Its density was measured in a methyl salicylate/carbon tetrachloride liquid mixture and was found equal to 1.41 g/cm$^3$.

**Preparation of the Solutions of the BSA/Polyelectrolyte Mixtures.** The solutions used for SLS and DLS measurements were prepared by mixing parent dilute solutions of BSA and of the two graft copolymers, G75 and G87.5, of a concentration of about 1.00 x 10$^{-3}$ g/cm$^3$, prepared in 0.05 M citric acid-phosphate buffer, in NaCl 0.1 M, under gentle agitation for 24 hours. Solutions used for SANS measurements were prepared by mixing equal volumes of parent solutions 1.0 x 10$^{-2}$ g/cm$^3$ BSA, 1.04 x 10$^{-2}$ G75 and 2.12 x 10$^{-2}$ G87.5 in 0.15 M citric acid-phosphate buffer in $D_2O$.

## Results and Discussion

Figure 1 shows the variation of the light scattering intensity, measured at 90$^o$, $I_{90}$, of two mixtures of BSA with G75 and G87.5, in dilute solution, versus pH. As pH decreases at values lower than 4.9, the pI of BSA, $I_{90}$ increases considerably, indicating the appearance of bigger particles as a result of complex formation between BSA and the two graft copolymers. In this pH range, BSA is positively charged and it is expected to form with the anionic AMPS units of the graft copolymers backbone, insoluble protein-polyelectrolyte complexes, due to Coulombic interaction.[2] The remaining, after the graft reaction, AA units, on the copolymer backbone, are not expected to play any role in this complexation, as they should be in their neutral form in this low pH region. Increase in the light scattering intensity reveals the formation of complexes soluble in water, due to the neutral PDMAM side chains of the two graft copolymers. We also observe that the increase in $I_{90}$ is less pronounced in the case of the BSA/G87.5 mixture, where the composition of the graft copolymer in PDMAM chains is higher. This behaviour should be attributed to the formation of less aggregated particles, as a result of the increased hydrophilic character of the complexes formed with G87.5 as it compares with G75. Moreover, complexation appears to weaken at pH lower than 3.25 – 3.5, in agreement also with previous turbidimetric and viscometric results.[19]

In order to clearly show the increase in light scattering intensity of the mixtures of BSA with the two graft polyelectrolytes with respect to the pure components and search for the stoichiometry of the complexes formed, we've measured the light scattering intensity of the mixtures BSA/G75 and BSA/G87.5 in dilute solution at pH = 3.0, in 0.1 M NaCl. Figure 2 shows the variation of the relative intensity, $I_{rel}$, of the two mixtures as a function of their weight fraction in BSA, $w_{BSA}$.

$I_{rel}$ is defined by the relation

$$I_{rel} = \frac{I}{I_{id}} \qquad (1)$$

where $I$ is the light scattering intensity at 90° and $I_{id}$ is the ideal intensity, which would be scattered by the same mixtures in the absence of any interaction between BSA and the graft copolymers, and it is calculated by the equation

$$I_{id} = w_1 I_1 + w_2 I_2 \qquad (2)$$

where $I_1$ and $I_2$ are the intensities of the two pure constituents and $w_1$, $w_2$ are their weight fractions in the mixture.

We observe that in both mixtures the relative light scattering intensity increases considerably, in comparison with that of the ideal behavior (dashed line), due to the complex formation between the anionic graft copolymers and the positively charged BSA. We also observe that $I_{rel}$ appears to have decreased considerably in the case of the BSA/G87.5 mixture, where the graft copolymer is of a higher composition in PDMAM side chains, as it compares with the G75, showing that as the composition of the graft copolymer in neutral PDMAM side chains increases, the formation of smaller complex particles is favored. We also observe that a maximum appears in a mixture composition dependent on the grafting degree of each copolymer. In the case of the mixture BSA/G75 the maximum appears at $w_{BSA}$ = 0.49, while for the mixture BSA /G87.5 it appears at $w_{BSA}$ = 0.32. These values indicate the stoichiometry of the complexes formed, and if we take into account that a complete neutralization of the oppositely charged substances (BSA and G75 or G87.5) occurs, then each BSA molecule should bear about 70 positive charges. Such a number seems to be plausible at pH = 3.0, as each BSA molecule contains 101 amine groups.[20]

The formation of graft polyelectrolyte – BSA complexes through Coulombic interactions was further confirmed by measuring the ζ-potential of the above G75/BSA and G87.5/BSA mixtures at different compositions, 0.25 < $w_{BSA}$ < 0.75. The values found were between -1 and +1 mV, showing that practically neutral complex particles are formed through charge neutralization.

Table 2 presents the SLS and DLS results obtained for the complexes formed between BSA and the two graft copolymers G75 and G87.5, at pH = 3.0, in their stoichiometric composition. According to the SLS results the complex particles formed between BSA and G75 present a larger radius of gyration than the particles formed between BSA and G87.5, equal to 30 nm, and a larger hydrodynamic radius, 83 nm as it compares with 25 and 65 nm, respectively. On the basis of the stoichiometry of these complexes and the molecular weights of the two graft copolymers (Table 1), approximately 14 BSA molecules should correspond to G75 and 13 to G87.5. The molar mass of such a "unit complex" would be 1.9 x $10^6$ Da and 2.7 x $10^6$ Da, respectively. According to these values and the apparent molecular weights of the two complexes obtained (Table 2), we can calculate their aggregation degrees equal to 2.2 and 1.5, respectively. Note that these aggregation numbers are averages, and necessarily influenced by the polydispersity in chain mass. This analysis shows that the aggregation is rather low, and that it is stronger in the case of the graft copolymer with the lower composition in PDMAM side chains that is the G75 copolymer, as a consequence of the lower hydrophilicity of the complex formed with BSA in this case. Anyway, it is remarkable that the complex nanoparticles formed between BSA and the two graft copolymers, G75 and G87.5, due to their high composition in neutral PDMAM side chains present a low aggregation, showing that it should not be so difficult to prepare complexes without any aggregation, through a better control of the involved parameters, such as grafting composition and probably the molecular weight of the graft copolymers. We consider that such an improvement could make these complexes more interesting candidates in drug delivery systems.

Figure 3(a) and (b) shows our SANS results, i.e. the variation of the neutron scattering intensity, *I*, versus the wave vector length, *q*, of the same polymer mixtures, BSA/G75 and BSA/G87.5 correspondingly, in solution in $D_2O$, at pH = 3.0. We observe that both mixtures show a considerable increase in the intensity scattered at low *q*, in comparison with the scattering of their two pure components. This increase is explained by the formation of dense hydrophobic complexes between the positively charged BSA molecules and the anionic backbone of the two graft copolymers, stabilized by their neutral hydrophilic PDMAM side chains. The values 3.1 and 2.9 obtained for the exponent d of the relation $I \sim q^{-d}$, in the intermediate wave vector length region 0.01 < q < 0.06, for the two mixtures, might indicate the fractal like structure of aggregates.[21, 22] The variation of *I* in the low *q* region is considered to represent the neutron scattering due to these protein/polyelectrolyte complexes formed. It corresponds, to a first approximation, to the Guinier regime of the scattering of individual noninteracting dense objects;[23] their radius leads to a characteristic decrease in *I*, whose magnitude is related to their mass. If the objects are spheres of radius *R*

$$I = I_o \exp(-R^2 q^2/5) \quad (3a)$$

with

$$I_o = \varphi(\Delta\rho)^2 V_c \quad (3b)$$

where $V_c$ denotes the volume of an individual object, *φ* the volume fraction of the objects, and $\Delta\rho$ the scattering contrast. Following equation (3a), the radii of the complexes of BSA with G75 and G87.5 are determined equal to 27 and 17 nm, respectively. These values should be compared with the values of 30 and 25 nm, obtained for the radii of gyration, and 83 and 65 nm, obtained for the hydrodynamic radii, respectively (Table 2). Here, we should point out that the values obtained by SLS for the radius of gyration and by DLS for the hydrodynamic radius characterize the whole complex particle, while the values obtained by SANS characterize their dense complex core. Also, according to the above results, the colloidal nanoparticles formed between BSA and the anionic graft copolymers are definitely smaller when the composition of the graft copolymer in PDMAM side chains is higher.

Moreover, the intensity values at q → 0, $I_o$, have been estimated to be equal to 70 and 26 cm$^{-1}$ for the two complexes of BSA with G75 and G87.5, respectively. Taking into account that these complexes are formed between the BSA molecules and the anionic backbone of the graft copolymers, their volume fraction $\varphi$ is estimated to be equal to 4.4 x 10$^{-3}$ in both cases. The scattering contrast, 4.4 x 10$^{10}$ cm$^{-2}$, was calculated by the equation $\Delta\rho = \rho_s - \rho_c$, where $\rho_s$ is the scattering length density of the solvent (D$_2$O), 6.4 x 10$^{10}$ cm$^{-2}$, and $\rho_c$ that of the dry complex, 2.0 x 10$^{10}$ cm$^{-2}$, calculated by means of the equation

$$\rho_c = \frac{\sum n_i l_i}{M} N_A d_c \qquad (4)$$

$l_i$ been the scattering length of each atom, $n_i$ the number of atoms in the particle, M its molecular weight, $N_A$ the number of Avogadro and $d_c$ its density. At this point we should emphasize that, for this calculation, we have taken into account the analytical composition of BSA.[24] Following the above obtained values for $I_o$, the values for the dry radii, $R_{dry}$, of the complexes of BSA with G75 and G87.5, equal to 12.5 and 9.0 nm, respectively, were obtained. From $R_{dry}$ we can obtain the molecular mass, $M_c$, of the dense core of the particle through[25]

$$M_c = (4/3)\pi R_{dry}^3 d_c N_A \qquad (5).$$

The values 7.0 x 10$^6$ and 2.6 x 10$^6$ Da for the two complexes, respectively, were obtained. These values should be compared with the values of 2.6 x 10$^6$ and 1.6 x 10$^6$ Da, which were calculated for the dense core of each complex from the SLS results. We see that the values found by SANS are higher, but they also seem to converge to the SLS values, as the content of the graft copolymer in hydrophilic PDMAM side chains increases. We attribute this discrepancy to the polydispersity of the aggregates and inhomogeneities in the complexes,[22] possibly related with the expected random distribution of the PDMAM side chains in the graft copolymers. Such an explanation seems to be justified also by the fact that this discrepancy is broader in case of the lower PDMAM composition sample, G75, where the hydrophobic-hydrophilic balance is certainly displaced to the hydrophobic side. Moreover, by calculating the volume of the dry particles we could find that the colloidal nanoparticles formed are 90 and 85% hydrated, respectively.

Finally, the particles formed should be comprised by a hydrophobic core of an insoluble complex of BSA with the anionic backbone of the graft copolymer, and a hydrophilic corona, of neutral PDMAM side chains. A similar core-shell structure has been also proposed and adequately described for colloids made of neutral/polyelectrolyte diblock copolymers and oppositely charged surfactants.[26] Considering the BSA molecule as an ellipsoidal revolution [27] a representation of the complex formed could be that of Scheme 2. According to this schematic depiction, the BSA molecules are enclosed in the insoluble BSA-polyelectrolyte complex, the core of a colloidal nanoparticle, stabilized by a hydrophilic corona, which is comprised by the neutral PDMAM side chains of the graft copolymer. The stability of these complexes at low pH, makes them suitable as protein careers in oral drug delivery systems. They could protect the protein in the low pH gastric environment and release it in the higher pH intestine environment.

**Conclusions**

In the present work, we studied and characterized colloidal nanoparticles formed through Coulombic interactions and charge neutralization between BSA and two anionic graft copolymers of poly (sodium acrylate-*co*-sodium 2-acrylamido-2-methyl-1-propanesulphonate) (P(Na-*co*-NaAMPS)) grafted with neutral PDMAM chains, G75 and G87.5, containing 75 and 87.5 wt. % of PDMAM respectively, in dilute solution, in citric acid-phosphate buffers, in 0.1 M NaCl, at pH 3.0. Core-corona type nanoparticles are formed, comprised by a hydrophobic BSA/P(Na-*co*-NaAMPS), protein/polyelectrolyte, complex core, with a radius of 26 or 18 nm, surrounded by a hydrophilic corona of PDMAM neutral chains, and presenting a hydrodynamic radius of 83 or 65 nm, respectively. The nanoparticles formed present low aggregation numbers, between 2 and 6, while they tend to be molecularly dispersed as the composition of the graft copolymers in neutral PDMAM side chains increases. Finally, we consider that such systems could be proved useful in oral drug delivery systems, where the gastrointestinal tract should be used.

**Acknowledgement.** This research project has been supported by the European Commission under the 6th Framework Programme through the Key Action: Strengthening the European Research Area, Research Infrastructures. Contract n°: HII3-CT-2003-505925.

**Table 1.** Graft Copolymers Synthesized and Used in this Study

| graft copolymer | wt. composition in PDMAM | $M_w \times 10^{-6}$ |
|---|---|---|
| G75 | 75 % | 0.95 |
| G87.5 | 87.5 % | 1.8 |

**Table 2.** Compexes Characterization, SLS and DLS results

| complex | $M_{app}$ x $10^{-6}$, Da | $R_G$, nm | $R_H$, nm |
|---|---|---|---|
| BSA/G75 | 4.1 | 30 | 83 |
| BSA/G87.5 | 4.0 | 25 | 65 |

**Scheme 1.** A schematic depiction of the anionic graft copolymers

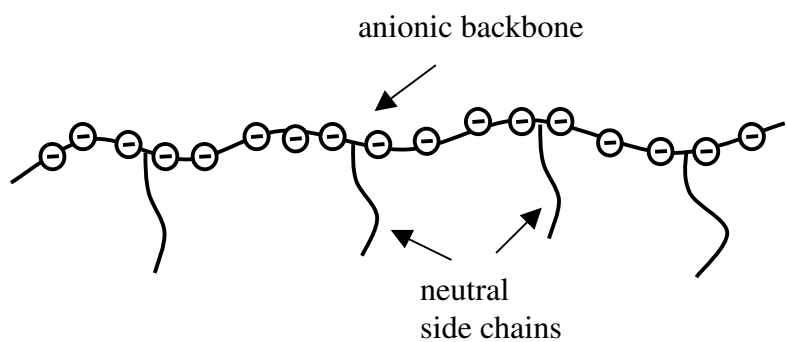

**Scheme 2.** A Schematic Depiction of the Colloidal Core-Corona Nanoparticles Formed Between BSA and G75 (or G87.5) at pH = 3.0.

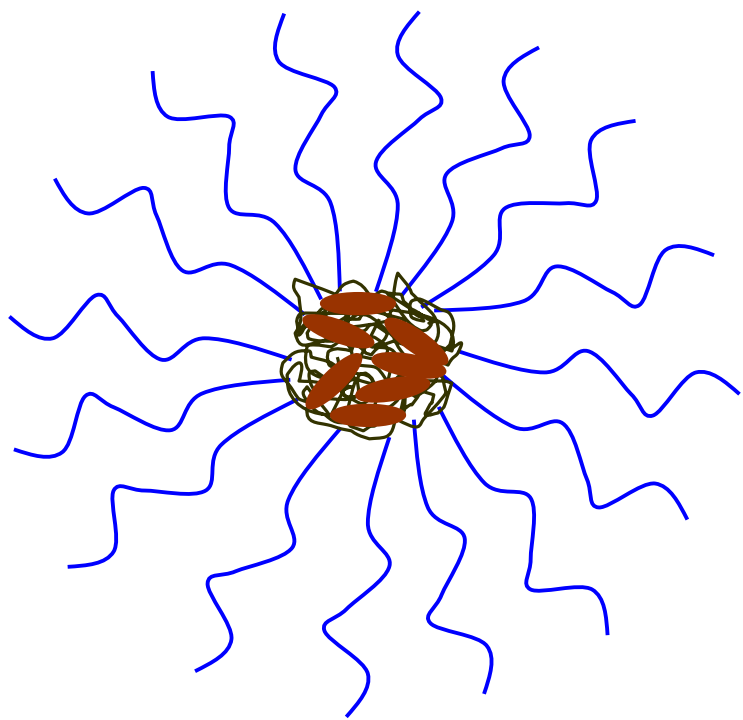

Figure Captions

**Figure 1**. Variation of the light scattering intensity, measured at 90°, $I_{90}$, as a function of pH, for the BSA mixtures with G75, (●) and G87.5, (■) in dilute solution, in 0.05 M citric acid - phosphate buffers, in 0.1 M NaCl. The concentrations were: BSA, 5.0 x $10^{-4}$ g/cm$^3$; G75, 4.0 x $10^{-4}$ g/cm$^3$; and G87.5, 8.0 x $10^{-4}$ g/cm$^3$.

**Figure 2**. Variation of the light scattering relative intensity, $I_{rel}$, of the BSA/G75, (●) and BSA/G87.5, (■) mixtures, in dilute solution, at pH = 3.0, in NaCl 0.1 M, versus the weight fraction in BSA, $w_{BSA}$. The concentrations of the pure initial solutions used were: BSA, 5.0 x $10^{-4}$ g/cm$^3$; G75, 4.0 x $10^{-4}$ g/cm$^3$; and G87.5, 8.0 x $10^{-4}$ g/cm$^3$.

**Figure 3**. (a) Variation of the neutron scattering intensity, $I$, versus the wave vector, $q$, for the BSA/G75 mixture, in a 0.15 M citric acid - phosphate buffer, dilute solution in D$_2$O, at the charge neutralization stoichiometric composition, $w_{BSA}$ = 0.49 (■) and for pure BSA (●) and pure G75 (○) solutions, at pH = 3.0; dash line represents the slope at intermediate $q$, and solid line represents the Guinier fitting, Eq. (3a).
(b) As above, for the BSA/G87.5 mixture, $w_{BSA}$ = 0.32 (■); pure BSA (●); and pure G75 (○) solutions.
The concentrations were: BSA, 5.0 x $10^{-3}$ g/cm$^3$; G75, 5.02 x $10^{-3}$ g/cm$^3$; and G87.5, 1.06 x $10^{-2}$ g/cm$^3$.

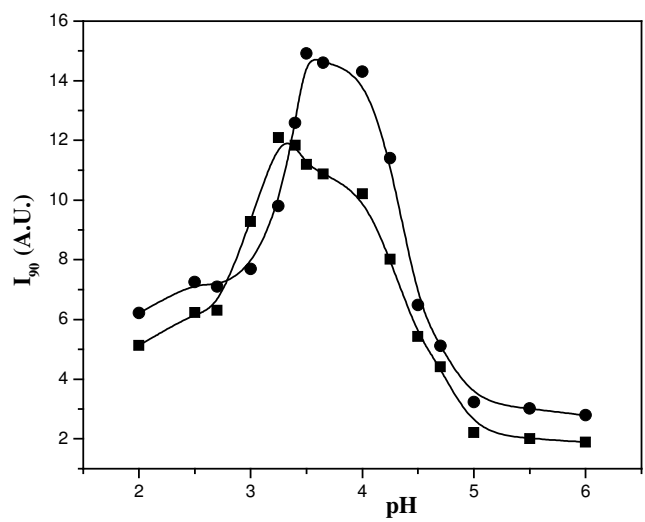

Figure 1

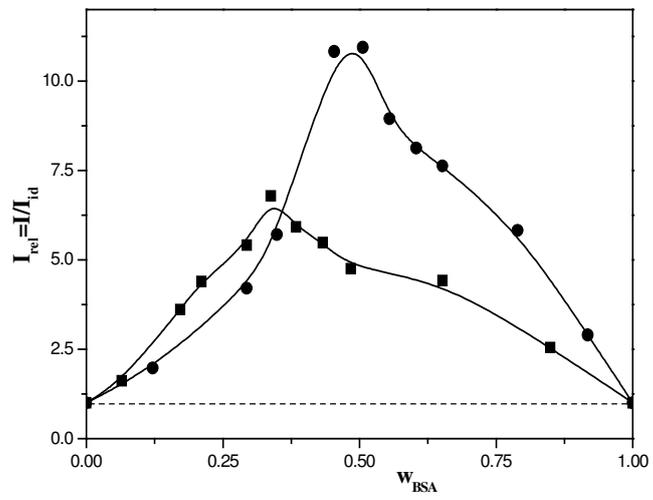

Figure 2

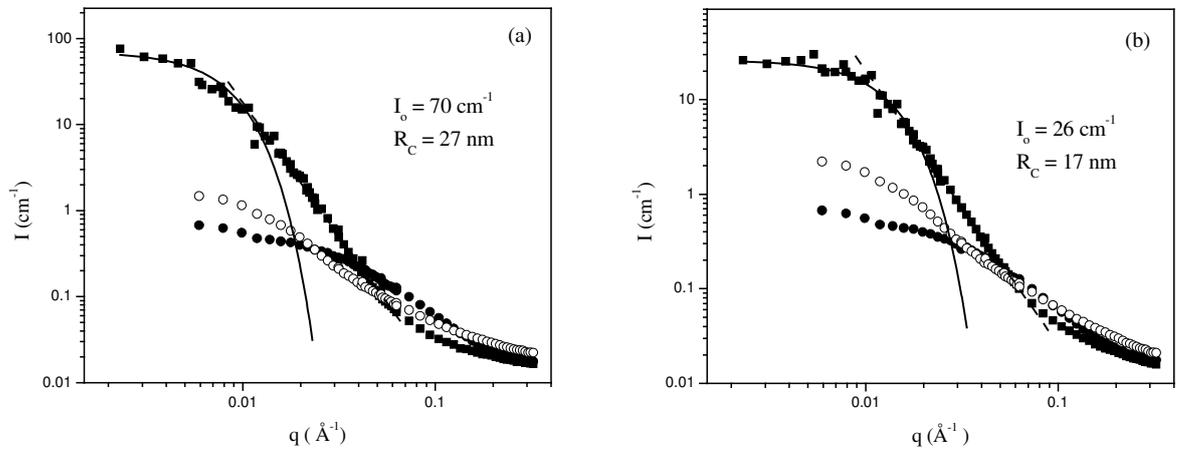

Figure 3

Table of Contents Graphic

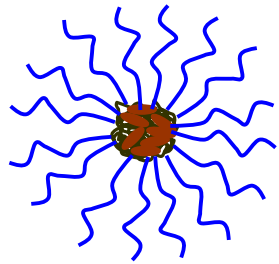